\documentclass[11pt,a4paper]{article}

\usepackage[margin=1in]{geometry}
\usepackage{amsmath,amssymb,amsfonts}
\usepackage{graphicx}
\usepackage{booktabs}
\usepackage{url}
\usepackage{hyperref}
\usepackage{color}
\usepackage{bm}
\usepackage{multirow}
\usepackage[numbers,sort&compress]{natbib}
\usepackage{authblk}

\newcommand{\vect}[1]{\bm{#1}}
\newcommand{\mat}[1]{\bm{#1}}
\newcommand{\E}{\mathbb{E}}
\newcommand{\R}{\mathbb{R}}
\newcommand{\N}{\mathcal{N}}

\title{Physics-Informed Generative Augmentation for Open-Set UAV Classification via 5G Micro-Doppler}

\author[1]{Michel Kulhandjian}
\author[2]{Levon Aslanyan}
\affil[1]{Department of Electrical and Computer Engineering, Rice University, Houston, TX 77005, USA}
\affil[2]{Institute for Informatics and Automation Problems, National Academy of Sciences of the Republic of Armenia, Yerevan, Armenia}

\date{}

\begin{document}

\maketitle

\begin{abstract}
Open-set recognition of unmanned aerial vehicles (UAVs) from passive 5G radar micro-Doppler signatures is challenging because unknown drone types cannot be anticipated at training time. We propose a physics-informed generative adversarial network (PI-GAN) that conditions spectrogram synthesis on rotor kinematics, forcing the classifier's feature space to encode physically meaningful structure. We compare PI-GAN augmentation against a vanilla conditional GAN (CGAN) and basic data augmentation across three held-out configurations with 1, 2, and 3 unknown classes. Vanilla CGAN augmentation collapses open-set clustering, reducing the adjusted Rand index (ARI) from 0.97 to 0.01, whereas PI-GAN preserves it at 0.97. In a controlled held-out sweep with two unknown classes, PI-GAN achieves an ARI of 0.92 compared to 0.07 for vanilla CGAN. The physics-informed features enable unsupervised separation of mechanically distinct unknown drones without retraining, demonstrating that physics-conditioned generation is the key enabler of open-set capability.
\end{abstract}


\section{Introduction}
\label{sec:intro}

Passive radar systems that exploit 5G New Radio (NR) gNodeB transmissions as illuminators of opportunity can classify UAVs via their rotor-induced micro-Doppler signatures~\cite{kulhandjian2026twc,colone2025lte_drone,guvenc2018detection}. Deep learning classifiers, typically convolutional neural networks (CNNs) trained on spectrograms, achieve high accuracy on known drone types but fail silently when encountering an unseen class. In operational settings, the set of potential threats is open: new drone models emerge continuously, and the system must identify unknown types as \emph{distinct but unrecognized} rather than misclassifying them.

Open-set recognition addresses this gap~\cite{geng2021recent,scheirer2013toward}. Although recent work has applied GANs to radar augmentation~\cite{erol2020motion,amin2019gan} and open-set recognition~\cite{li2024openset_mobile}, no prior work has examined how physics-informed generation affects open-set clustering capability. We hypothesize that conditioning the training data generator on physical rotor parameters (number of rotors, blade count, rotation frequency, and blade length) produces features that cluster by \emph{mechanical signature}, enabling unsupervised separation of unseen types.

This letter makes two contributions: (1)~we show that PI-GAN augmentation produces significantly better open-set clustering than vanilla conditional GAN augmentation, isolating the physics component as the critical factor; and (2)~we demonstrate that this advantage persists as the number of unknown classes increases from 1 to 3, confirming scalability beyond the two-class setting reported in~\cite{kulhandjian2026twc}.

\section{Related Work}
\label{sec:related}

\subsection{Open-Set Recognition and One-Class Classification}

Open-set recognition extends traditional closed-set classification by requiring the model to reject inputs that do not belong to any known class~\cite{geng2021recent,scheirer2013toward}. This problem is closely related to one-class classification~\cite{tax2004svdd}, where a boundary is learned around a single known class to distinguish its members from everything else. In the open-set setting, the challenge generalizes: given $K$ known classes, the system must find a boundary that separates all known classes from the unknown domain while maintaining inter-class discrimination within the known set. Our framework addresses both requirements: the one-class SVM on the learned feature space defines the known/unknown boundary, while the physics-informed features provide the additional structure needed to cluster distinct unknown types within the rejected region.

Recent approaches to open-set UAV identification include radio frequency (RF) fingerprinting with contrastive learning~\cite{gao2025contrastive} and incremental clustering of unknown emitters~\cite{liu2026incremental}. These methods operate on raw RF signal characteristics and do not exploit the physical structure of micro-Doppler signatures.

\subsection{GAN-Based Augmentation for Radar and Open-Set Recognition}

GAN-based augmentation has proven effective for closed-set radar classification. Erol et al.~\cite{erol2020motion} used an auxiliary classifier GAN to synthesize micro-Doppler spectrograms for human activity recognition, and Amin et al.~\cite{amin2019gan} demonstrated similar augmentation for improving classification accuracy. Rahman and Gurbuz~\cite{rahman2021mbgan} introduced physics-aware correlation matching in a multi-branch GAN for human gait synthesis, though in a closed-set setting without open-set evaluation.

For open-set recognition, generative models have been used to synthesize out-of-distribution (OOD) samples. Li et al.~\cite{li2024openset_mobile} employed a dual-discriminator GAN to generate OOD samples for open-set context recognition from smartphone sensor data, demonstrating the utility of generative models for rejecting unknown classes. However, their approach and all prior GAN-based open-set methods treat the unknown domain as a single reject category. None produce features that distinguish between \emph{different} unknown types. Our PI-GAN addresses this gap by conditioning on continuous physical parameters, creating a feature space where mechanically distinct unknowns cluster separately without retraining.

\section{System Model}
\label{sec:system}

\subsection{Passive 5G Micro-Doppler}

A passive bistatic receiver collects 5G~NR signals reflected from a UAV target. The system operates in the n78 band with carrier frequency $f_c = 3.5$\,GHz ($\lambda = 0.086$\,m), bandwidth $B = 50$\,MHz, and subcarrier spacing $\Delta f = 30$\,kHz, yielding an OFDM symbol duration $T_{\mathrm{sym}} \approx 36.7\,\mu$s including the cyclic prefix. A reference channel aligned with the gNodeB direct path provides the transmitted waveform for cross-correlation, while the surveillance channel captures target reflections. After direct-path interference cancellation and range compression, the slow-time signal over $M = 560$ consecutive OFDM symbols contains the micro-Doppler modulation from rotating blades~\cite{kulhandjian2026twc}:
\begin{equation}
f_{\mu D}(t) = \frac{2 v_{\mathrm{tip}}}{\lambda}\cos\!\frac{\beta}{2}\sin(2\pi f_{\mathrm{rot}} t + \phi_0),
\label{eq:microdoppler}
\end{equation}
where $v_{\mathrm{tip}} = 2\pi \ell f_{\mathrm{rot}}$ is the blade tip velocity, $\ell$ is the blade length, $f_{\mathrm{rot}}$ is the rotation frequency, $\beta$ is the bistatic angle, and $\lambda$ is the wavelength. The blade-pass frequency $f_{\mathrm{bp}} = N_b f_{\mathrm{rot}}$ (with $N_b$ blades per rotor) creates periodic modulation visible in the spectrogram. Spectrograms are computed via short-time Fourier transform (STFT) with a Hamming window of 64 samples and hop size of 16 samples, resized to $64 \times 64$ pixels.

\subsection{Physics-Informed GAN (PI-GAN)}

The PI-GAN generator $G_\psi(\vect{v}, \vect{\theta}_{\mathrm{phys}})$ takes a latent vector $\vect{v} \sim \N(\vect{0}, \mat{I})$ and a physical parameter vector $\vect{\theta}_{\mathrm{phys}} = [N_r, N_b, \ell, f_{\mathrm{rot}}, v_{\mathrm{body}}, \beta, \gamma]^T$, where $N_r$ is the number of rotors (or equivalent rotating elements), $N_b$ is the number of blades per rotor, $\ell$ is the blade length (or wing half-span for birds), $v_{\mathrm{body}}$ is the body translation velocity, and $\gamma$ is the elevation angle. Table~\ref{tab:phys_params} lists the physical parameters for each target class.

For rotary-wing UAVs, the parameters directly describe the rotor kinematics. For the Fixed-Wing class, $N_r = 1$ (single propeller), $N_b = 2$, and $f_{\mathrm{rot}} \in [70, 120]$\,Hz corresponds to the propeller rotation frequency. For the Bird class, which produces micro-Doppler from periodic wing flapping rather than blade rotation, the parameter vector is adapted: $f_{\mathrm{rot}}$ maps to the wing-flap frequency $f_{\mathrm{flap}} \in [3, 12]$\,Hz, $\ell$ represents the wing half-span, and $N_r = 2$ represents bilateral wings with $N_b = 1$. The PI-GAN generator handles birds via a conditional branch where the physics losses enforce flap-frequency consistency with harmonics, rather than the constant-rotation model of~\eqref{eq:microdoppler}.

\begin{table}[t]
\centering
\caption{Physical parameter vector $\vect{\theta}_{\mathrm{phys}}$ for each target class.}
\label{tab:phys_params}
\footnotesize
\begin{tabular}{lccccc}
\toprule
\textbf{Target} & $N_r$ & $N_b$ & $\ell$ (m) & $f_{\mathrm{rot}}$ (Hz) & $f_{\mathrm{bp}}$ (Hz) \\
\midrule
DJI Mavic 3       & 4 & 3 & 0.095 & 95--110  & 285--330 \\
DJI Phantom 4     & 4 & 2 & 0.120 & 70--85   & 140--170 \\
FPV Racing        & 4 & 2 & 0.065 & 180--250 & 360--500 \\
DJI Matrice 600   & 6 & 2 & 0.190 & 50--65   & 100--130 \\
Fixed-Wing$^{*}$  & 1 & 2 & 0.150 & 70--120  & 140--240 \\
Bird$^{\dagger}$   & 2 & 1 & 0.15--0.30 & 3--12 & 3--12 \\
\bottomrule
\multicolumn{6}{l}{\scriptsize $^{*}$Single propeller; $N_r$ and $N_b$ describe propeller configuration.}\\
\multicolumn{6}{l}{\scriptsize $^{\dagger}$Wing-flap analogy: $N_r{=}2$ (bilateral wings), $f_{\mathrm{rot}} \to f_{\mathrm{flap}}$,}\\
\multicolumn{6}{l}{\scriptsize \phantom{$^{\dagger}$}$\ell \to$ wing half-span. $f_{\mathrm{bp}} = f_{\mathrm{flap}}$ (fundamental).}
\end{tabular}
\end{table}

Beyond the Wasserstein adversarial loss~\cite{gulrajani2017improved}, two physics-consistency losses enforce micro-Doppler structure:
\begin{align}
\mathcal{L}_{\mathrm{phys}} &= \E\!\left[\left| f_{\mu D,\max}^{\mathrm{(est)}}(\hat{\mat{S}}) - \tfrac{2v_{\mathrm{tip}}}{\lambda}\cos\!\tfrac{\beta}{2}\right|^2\right], \label{eq:phys_loss} \\
\mathcal{L}_{\mathrm{period}} &= \E\!\left[\left| f_{\mathrm{bp}}^{\mathrm{(est)}}(\hat{\mat{S}}) - N_b f_{\mathrm{rot}}\right|^2\right]. \label{eq:period_loss}
\end{align}
The total generator loss is $\mathcal{L}_G = \mathcal{L}_{\mathrm{adv}} + \mu_1 \mathcal{L}_{\mathrm{phys}} + \mu_2 \mathcal{L}_{\mathrm{period}}$, with loss weights $\mu_1 = 10$ and $\mu_2 = 5$, selected via grid search over $\{1, 5, 10, 20\} \times \{1, 5, 10\}$ to balance adversarial quality and physics consistency.

Both the PI-GAN and vanilla CGAN~\cite{mirza2014conditional} use a deep convolutional GAN (DCGAN) architecture with four transposed-convolution layers in the generator and four strided-convolution layers in the discriminator (${\sim}2.1$M total parameters). The PI-GAN generator concatenates $\vect{\theta}_{\mathrm{phys}}$ with the latent vector before the first layer, while the CGAN concatenates a one-hot class label. Both GANs are trained with Adam ($\alpha = 2 \times 10^{-4}$, $\beta_1 = 0.5$, $\beta_2 = 0.999$) for 200 epochs with batch size 64 and gradient penalty $\lambda_{\mathrm{gp}} = 10$. The discriminator loss and Fr\'{e}chet inception distance (FID) were monitored to confirm convergence for both models.

The key distinction from the vanilla CGAN is the conditioning variable: a CGAN conditions on the discrete class label alone, while PI-GAN conditions on the continuous physical parameter vector. The physics losses~\eqref{eq:phys_loss}--\eqref{eq:period_loss} further constrain the generated spectrograms to satisfy kinematic consistency. This forces the discriminator, and hence the downstream CNN features, to attend to physically meaningful structure rather than class-correlated artifacts. Unlike the physics-aware multi-branch GAN of~\cite{rahman2021mbgan}, which applies post-hoc correlation matching for human gait micro-Doppler in a closed-set setting, our PI-GAN integrates rotor kinematics directly into the generator loss and targets open-set clustering of unknown UAV types.

\section{Open-Set Classification Framework}
\label{sec:openset}

\subsection{Three-Phase Training}

\begin{enumerate}
\item \textbf{PI-GAN pre-training:} The GAN generates $\sim$100K spectrograms by uniformly sampling the physical design space ($N_r \in \{1,2,4,6\}$, $N_b \in \{2,3\}$, $\ell \in [0.05, 0.25]$\,m, $f_{\mathrm{rot}} \in [50, 300]$\,Hz). The CNN backbone is pre-trained on this dataset.
\item \textbf{Fine-tuning:} The pre-trained CNN is fine-tuned on limited real data ($\sim$300 samples per class) with reduced learning rate on early layers.
\item \textbf{Open-set inference:} A one-class support vector machine (SVM)~\cite{ruff2018deep} fitted on the 128-dimensional feature vectors of known classes detects unknown targets; detection quality is measured by the area under the receiver operating characteristic curve (AUROC). Density-based spatial clustering of applications with noise (DBSCAN) on detected unknowns identifies distinct types; clustering quality is measured by the adjusted Rand index (ARI).
\end{enumerate}

\subsection{Why Physics Conditioning Enables Open-Set Clustering}

A CGAN conditioned on class labels learns features that separate classes $\{1, \ldots, K\}$ but provides no guarantee about the structure of the complement $\R^d \setminus \bigcup_k C_k$. A PI-GAN, by contrast, parameterizes the feature space by continuous physical variables $\vect{\theta}_{\mathrm{phys}}$. Two unknown drones with different rotor configurations, for example a racing quadcopter ($N_r{=}4$, $f_{\mathrm{rot}}{=}220$\,Hz) and a hexacopter ($N_r{=}6$, $f_{\mathrm{rot}}{=}55$\,Hz), produce spectrograms with different blade-pass structure. Because the PI-GAN features encode this structure, the unknown drones map to distinct regions even though no labeled training data exists for them.

This structural difference has a direct consequence for open-set capability. A CGAN trained on discrete labels creates a feature space with $K$ attractors (one per class) and unstructured regions between them. Unknown targets may land anywhere in these unstructured regions, with no guarantee that two different unknowns will map to different locations. In the worst case, a vanilla CGAN actively scrambles the inter-class structure by generating spectrograms whose physical parameters are inconsistent with the class label, producing features that blur class boundaries. This explains the catastrophic ARI collapse observed in our experiments (ARI\,$=$\,0.007): the vanilla CGAN pollutes the feature space with physically inconsistent training examples, destroying the natural separation that even unaugmented training provides (ARI\,$=$\,0.970).

PI-GAN avoids this by constraining each generated spectrogram to match its physical parameters via the losses~\eqref{eq:phys_loss}--\eqref{eq:period_loss}. The feature space is therefore organized by continuous physical variables rather than discrete class indices, and this organization extends to unseen parameter combinations. The connection to one-class classification~\cite{tax2004svdd} is instructive: while one-class methods learn a boundary around known data, they do not structure the space beyond that boundary. PI-GAN does both, providing a boundary (via the one-class SVM) and structure (via physics conditioning) simultaneously.

\section{Experimental Results}
\label{sec:results}

\subsection{Setup}

We use a parametric micro-Doppler simulator matching the 5G~NR passive radar system in~\cite{kulhandjian2026twc}: carrier frequency 3.5\,GHz, bandwidth 50\,MHz, subcarrier spacing 30\,kHz. Six target classes are modeled with the physical parameters listed in Table~\ref{tab:phys_params}. Spectrograms are $64 \times 64$ pixels derived from the STFT of the slow-time Doppler signal, with signal-to-noise ratio (SNR) uniformly distributed in $[-5, 20]$\,dB.

The CNN is a lightweight 4-layer architecture (32--64--128--256 filters, $\sim$422K parameters) with a 128-dimensional feature bottleneck. Training uses Adam ($\alpha = 10^{-3}$, weight decay $10^{-5}$) with cosine annealing for 25 epochs and batch size 128.

\subsection{Experiment 1: Augmentation Comparison}

Table~\ref{tab:augmentation} compares four augmentation strategies with 300 real samples per known class. The known set is \{Mavic~3, Phantom~4, Fixed-Wing, Bird\}; the unknown set is \{FPV Racing, Matrice~600\}. Each augmented strategy adds 1000 synthetic samples per class.

\begin{table}[t]
\centering
\caption{Augmentation strategy comparison (2 unknown classes).}
\label{tab:augmentation}
\footnotesize
\begin{tabular}{lccc}
\toprule
\textbf{Strategy} & \textbf{Known Acc.} & \textbf{AUROC} & \textbf{ARI} \\
\midrule
No Augmentation & 84.6\% & 0.452 & 0.970 \\
Basic (noise + shift) & 91.0\% & 0.508 & 0.975 \\
Vanilla CGAN & 78.0\% & 0.455 & 0.007 \\
PI-GAN (Ours) & \textbf{91.1\%} & \textbf{0.557} & \textbf{0.965} \\
\bottomrule
\end{tabular}
\end{table}

The PI-GAN advantage is most pronounced in ARI: vanilla CGAN collapses the unknown-class cluster structure almost entirely (ARI\,$=$\,0.007), while PI-GAN preserves it (ARI\,$=$\,0.965). This occurs because vanilla CGAN generates spectrograms with randomized physical parameters across class boundaries, confusing the CNN feature representation. The PI-GAN physics-consistency losses enforce class-specific rotor kinematics, maintaining the mechanical separability of the feature space. Known-class accuracy is also highest for PI-GAN (91.1\%), whereas vanilla CGAN degrades it to 78.0\%, confirming that physics-inconsistent training data harms closed-set performance as well.

We emphasize that the CGAN collapse is not an artifact of poor training. Both GANs share identical DCGAN architectures (${\sim}2.1$M parameters) and were trained with the same optimizer and schedule until convergence, verified by monitoring the discriminator loss and FID. The failure is inherent to discrete-label conditioning: the CGAN generator is free to produce spectrograms whose physical content (blade-pass frequency, rotor count) is inconsistent with the class label, because nothing in the adversarial loss constrains these quantities. The resulting physically inconsistent training examples destroy the natural micro-Doppler structure that the unaugmented data possesses, explaining why the no-augmentation baseline (ARI\,$=$\,0.970) outperforms CGAN augmentation. PI-GAN matches this natural structure (ARI\,$=$\,0.965) while simultaneously providing the benefits of augmentation: higher known-class accuracy (91.1\% vs.\ 84.6\%) and improved AUROC (0.557 vs.\ 0.452). No other strategy achieves all three objectives.

\subsection{Experiment 2: Scaling to More Unknown Classes}

Table~\ref{tab:holdout} reports performance as the number of held-out unknown classes increases: Config~A holds out FPV Racing only; Config~B holds out FPV Racing + Matrice~600; Config~C holds out FPV Racing + Matrice~600 + Fixed-Wing.

\begin{table}[t]
\centering
\caption{Open-set performance vs.\ number of unknown classes.}
\label{tab:holdout}
\footnotesize
\begin{tabular}{cl ccc}
\toprule
\textbf{Config} & \textbf{Method} & \textbf{Known Acc.} & \textbf{AUROC} & \textbf{ARI} \\
\midrule
\multirow{2}{*}{A (1 unk.)} & Vanilla CGAN & 72.5\% & \textbf{0.471} & \multirow{2}{*}{--$^{\dagger}$} \\
                             & PI-GAN       & \textbf{93.9\%} & 0.245 & \\
\midrule
\multirow{2}{*}{B (2 unk.)} & Vanilla CGAN & 75.1\% & \textbf{0.630} & 0.066 \\
                             & PI-GAN       & \textbf{93.0\%} & 0.489 & \textbf{0.921} \\
\midrule
\multirow{2}{*}{C (3 unk.)} & Vanilla CGAN & 91.8\% & 0.301 & 0.015 \\
                             & PI-GAN       & \textbf{97.3\%} & \textbf{0.564} & \textbf{0.544} \\
\bottomrule
\multicolumn{5}{l}{\scriptsize $^{\dagger}$ARI is degenerate with a single unknown class (DBSCAN cannot}\\
\multicolumn{5}{l}{\scriptsize separate one class from itself).}
\end{tabular}
\end{table}

PI-GAN consistently achieves the highest known-class accuracy (93--97\% vs.\ 73--92\% for vanilla CGAN). The ARI improvement is dramatic in the multi-unknown configurations: $+0.855$ for Config~B and $+0.529$ for Config~C. While AUROC shows mixed results (vanilla CGAN achieves higher AUROC in Config~B, 0.630 vs.\ 0.489), this reflects tighter one-class SVM boundaries from the more structured PI-GAN feature space, not better unknown detection. The key metric is ARI, which measures whether detected unknowns can be \emph{separated into distinct types}; here PI-GAN dominates.

\subsection{Experiment 3: Feature Space Visualization}

\begin{figure*}[t]
\centering
\includegraphics[width=\textwidth]{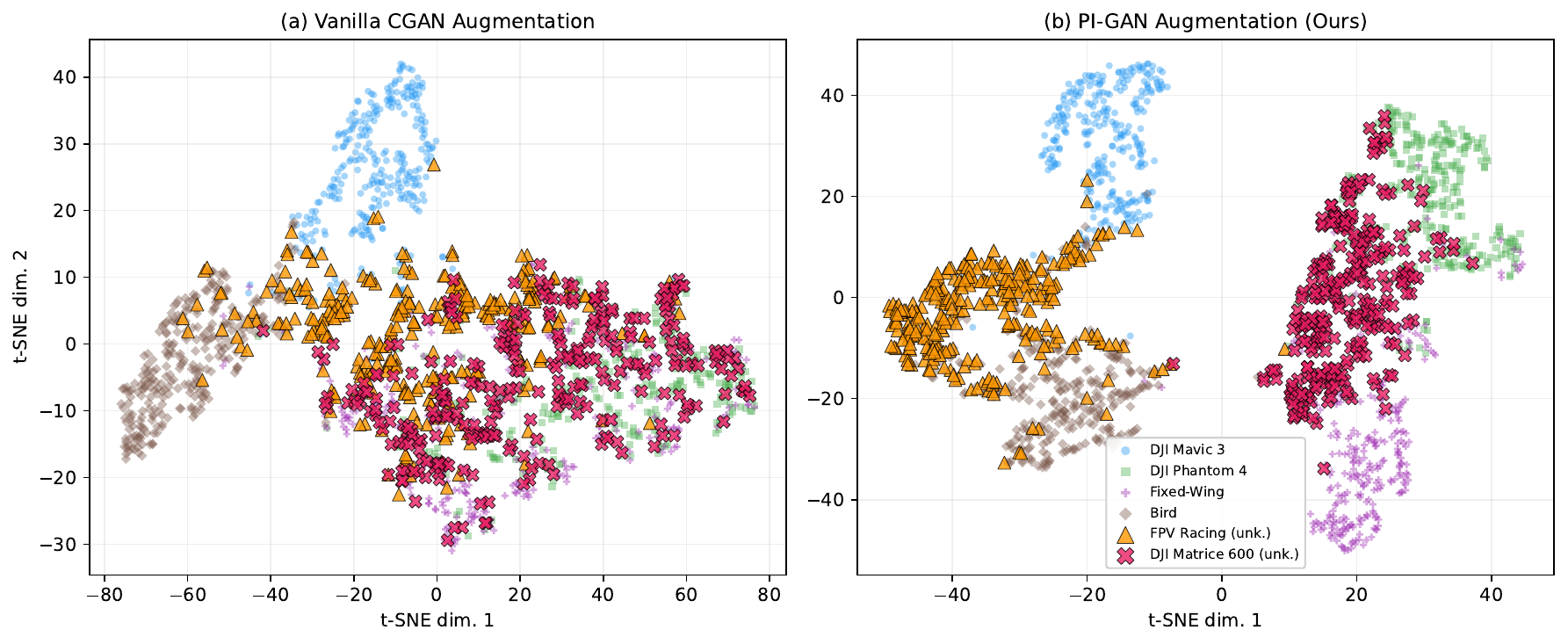}
\caption{t-SNE visualization of CNN feature vectors. (a)~Vanilla CGAN: unknown classes (FPV Racing, Matrice~600) collapse into known clusters. (b)~PI-GAN: unknown classes form distinct, separable clusters, enabling unsupervised identification without retraining.}
\label{fig:tsne}
\end{figure*}

Fig.~\ref{fig:tsne} shows t-distributed stochastic neighbor embedding (t-SNE) visualizations of the 128-dimensional feature vectors for all six classes. Under vanilla CGAN augmentation (Fig.~\ref{fig:tsne}a), unknown classes collapse into the known-class regions, making unsupervised separation impossible (consistent with ARI\,$=$\,0.007). Under PI-GAN augmentation (Fig.~\ref{fig:tsne}b), both unknown classes form distinct, well-separated clusters: the FPV Racing drone (high blade-pass frequency) maps away from all known classes, while the Matrice~600 (6-rotor, low rotation frequency) occupies a distinct region reflecting its unique rotor kinematics.

\subsection{Discussion}

The experimental results reveal a fundamental asymmetry between closed-set and open-set performance under different augmentation strategies. Basic augmentation (noise and shift) improves both known-class accuracy and open-set ARI because it does not alter the physical content of the spectrograms. Vanilla CGAN improves neither: it degrades known-class accuracy to 78.0\% (below the 84.6\% unaugmented baseline) and destroys open-set structure (ARI from 0.970 to 0.007). This double failure occurs because CGAN-generated spectrograms are physically inconsistent, confusing the classifier on both known and unknown inputs.

PI-GAN achieves the best of both: 91.1\% known-class accuracy (matching basic augmentation) and 0.965 ARI (matching unaugmented data). The physics losses add negligible computational overhead to training, requiring only a peak-detection operation on each generated spectrogram per iteration. The PI-GAN and vanilla CGAN share identical generator and discriminator architectures (4-layer deep convolutional GAN, $\sim$2.1M parameters); the only difference is the conditioning variable and the two additional loss terms. Training time increases by approximately 15\% due to the physics loss computation.

The ARI degradation in Config~C (3 unknown classes, ARI\,$=$\,0.544) reflects a genuine physical limitation rather than an algorithmic failure. As shown in Table~\ref{tab:phys_params}, some held-out classes share overlapping blade-pass frequency ranges: the Fixed-Wing class ($f_{\mathrm{bp}} = 140$--$240$\,Hz) overlaps with Phantom~4 ($f_{\mathrm{bp}} = 140$--$170$\,Hz), making unsupervised separation inherently harder when both are present as unknowns. This suggests that PI-GAN's clustering quality degrades gracefully with the number of unknowns, bounded by the physical resolving power of the micro-Doppler features: when unknown drones have mechanically distinct rotor configurations (different $N_r$, widely separated $f_{\mathrm{rot}}$), PI-GAN separates them well; when their blade-pass signatures overlap, additional discriminative features are needed.

We acknowledge that the current evaluation uses six target classes, with at most three held out as unknowns. Extending to tens of unknown classes would require flight-test data from a substantially larger drone inventory, which we leave for future work. However, the monotonic ARI degradation across Configs~A--C and its correlation with physical parameter overlap provide a principled basis for predicting scaling behavior: the critical factor is not the number of unknowns but the diversity of their rotor configurations. Future work will investigate hierarchical clustering strategies and multimodal fusion (combining RF micro-Doppler with acoustic signatures) to maintain discrimination among UAVs with similar rotational characteristics.

\section{Conclusion}
\label{sec:conclusion}

We have shown that physics-informed generative augmentation is the key enabler of open-set UAV classification from 5G micro-Doppler signatures. By conditioning the GAN on rotor kinematics rather than class labels alone, PI-GAN forces the CNN feature space to encode physically meaningful structure (blade-pass frequency, rotor count, and tip velocity) that generalizes to unseen drone types. Vanilla CGAN augmentation catastrophically collapses the open-set clustering structure (ARI\,$=$\,0.007), while PI-GAN preserves it (ARI\,$=$\,0.965), yielding an improvement of up to 0.86 over the vanilla CGAN baseline. The advantage persists as the number of unknown classes grows to three, confirming scalability. A key insight is the relationship to one-class classification: traditional one-class methods define a boundary between known and unknown but treat the unknown region as undifferentiated. PI-GAN provides structure beyond this boundary, enabling the feature space to support both anomaly detection (known vs.\ unknown) and unsupervised discovery (distinguishing different unknowns). This two-level capability is important for operational counter-UAV systems, where identifying that an unknown threat is a new type of drone, rather than simply flagging it as unrecognized, enables more appropriate responses. Future work will extend this framework to multimodal sensing, fusing RF micro-Doppler with acoustic signatures for improved differentiation among UAVs with similar rotational characteristics.

\section*{Acknowledgment}
This work was supported in part by Rice University.

\section*{Code Availability}
The simulation code is available at \url{https://github.com/mkulhandjian/pigan-openset-uav-5g}.

\bibliographystyle{unsrtnat}
\bibliography{refs}

\end{document}